\documentclass[prl,aps,twocolumn,showpacs,floatfix]{revtex4}%
\usepackage{amsmath}
\usepackage{graphicx}
\usepackage{amsfonts}
\usepackage{amssymb}

\begin{document}

\title{Ring-shaped luminescence patterns in a locally photoexcited electron-hole bilayer}
\author{A. V. Paraskevov$^{1,2}$, S. E. Savel'ev$^{1}$}

\affiliation{$^1$Department of Physics, Loughborough University,
Loughborough LE11 3TU, United Kingdom\\ $^2$ Kurchatov Institute,
Moscow 123182, Russia}

\begin{abstract}

We report the results of molecular dynamics simulation of a
spatiotemporal evolution of the locally photoexcited electrons and
holes localized in two separate layers. It is shown that the
ring-shaped spatial pattern of luminescence forms due to the strong
in-layer Coulomb interaction at high photoexcitation power. In
addition, the results predict (i) stationary spatial oscillations of
the electron density in quasi one-dimensional case and (ii)
dynamical phase transition in the expansion of two-dimensional
electron cloud when threshold electron concentration is reached. A
possible reason of the oscillations and a theoretical interpretation
of the transition are suggested.

\end{abstract}


\maketitle

Exciton systems in semiconductor nanostructures can provide an
unique opportunity to obtain nonequilibrium superfluid transition in
solids \cite{Mos, KK, Loz, ex_sup}. Indeed, a remarkable phenomenon
has been experimentally discovered in the system of interwell
excitons in GaAs/AlGaAs double quantum wells (QWs) \cite{But_ring}:
local off-resonant photoexcitation of electrons (e) and holes (h)
gives rise to a macroscopic ring-shaped pattern of the luminescence
spatial distribution when the excitation power is above some
threshold value. The similar effects were observed later in other
systems \cite{snoke_nat, Tim, stern}.

The first theoretical explanation of the phenomenon was suggested
\cite{BLS, theor, Ha} within the diffusive transport model (DTM).
However, the model is not applicable to the experiments
\cite{But_ring} for the derivation of the ring radius \cite{com}. In
particular, the ring radius $R$ is derived within the DTM as
\begin{equation}
R=\lambda\exp\left(  -2\pi D_{e}n_{0}/P_{ex}\right)  , \label{0}%
\end{equation}
where $D_{e}$ is the diffusion coefficient for electrons, $n_{0}$ is
the equilibrium two-dimensional (2D) electron density, $P_{ex}$ is
the stationary photoexcitation power, and the electron depletion
length $\lambda\gg R$ \cite{BLS}. According to the model \cite{BLS},
the diffusion of holes is the only reason why they move out of the
laser excitation spot. However, according Eq.(\ref{0}) the ring
radius does not depend on the hole diffusion coefficient $D_{h}$ (in
\cite{BLS} $D_{h}$ is in the exponent also), i.e. the ring forms
even if all holes are left into the excitation spot ($D_{h}=0$).
This clearly unphysical result is not a consequence of the limiting
case $\lambda\gg R$, but is valid in the general case \cite{Ha}.
Moreover, the DTM can not explain in principle the condition on the
excitation power, i.e. that the external ring appears only when the
power exceeds some critical value, $P_{ex}>\left(P_{ex}\right)
_{c}\approx250$ $\mu W$ \cite{But_ring}. So the development of a
different theory of the ring pattern formation is necessary.

Here we present both quasi one-dimensional (1D) and 2D simulations
of the spatiotemporal dynamics of optically generated electrons and
holes that are located in two coupled layers (Fig.1). We show that
due to (i) large in-layer Coulomb interaction in the pumping region
at high excitation intensities and (ii) essential difference in
mobilities and effective masses for electrons and holes in GaAs, a
macroscopic spatial separation of the electron and hole densities
appears naturally. In contrast to the previous theoretical studies
\cite{BLS,theor,Ha}, the in-plane distributions for electrons and
holes are found self-consistently. The distribution of holes is
located much closer to the excitation spot than the electron one,
and their narrow overlap gives large ring-shaped luminescence
pattern. We also find that the dependence of exciton formation rate
on the electron-hole relative velocity is crucial for the
ring-shaped pattern formation. More surprisingly, we have clearly
observed stationary spatial oscillations of the electron density
outside the luminescence ring in 1D case. A possible origin of the
oscillations is a kinetic instability of two interacting electron
counter flows: faster flow from the center to the periphery and
slower back flow from the periphery to the luminescence region due
to the depletion of electron density there.

\begin{figure}[ptbh]
\includegraphics[width=8.8cm]{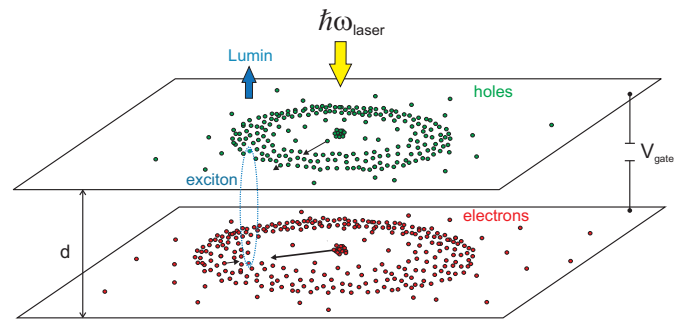}
\caption{\textbf{Schematic of the optically excited electron-hole
bilayer.} Both stationary laser pumping and luminescence are
perpendicular to the layers. The charge separation between the
layers is due to the external gate voltage $V_{gate}$. The
interlayer distance $d$ is essentially increased for the sake of
clarity.} \label{Fig1}
\end{figure}

\textit{Model.}---To describe the dynamics of $N$ hot electrons
and holes, we use the Newton equations of motion combined with the
exciton formation condition. The dimensionless equations read
\begin{align}
\ddot{\mathbf{r}}_{e}^{i}+\dot{\mathbf{r}}_{e}^{i} &  ={{\displaystyle\sum\limits_{j\neq i}}
}\frac{(\mathbf{r}_{e}^{i}-\mathbf{r}_{e}^{j})}{\left\vert \mathbf{r}_{e}%
^{i}-\mathbf{r}_{e}^{j}\right\vert ^{3}}-{{\displaystyle\sum\limits_{k}}
}\frac{\left(  \mathbf{r}_{e}^{i}-\mathbf{r}_{h}^{k}\right)  }{\left[  \left(
\mathbf{r}_{e}^{i}-\mathbf{r}_{h}^{k}\right)  ^{2}+d^{2}\right]  ^{3/2}%
},\label{e1}\\
\ddot{\mathbf{r}}_{h}^{i}+c_{1}\dot{\mathbf{r}}_{h}^{i} &  ={{\displaystyle\sum\limits_{j\neq i}}
}\frac{c_{2}(\mathbf{r}_{h}^{i}-\mathbf{r}_{h}^{j})}{\left\vert \mathbf{r}
_{h}^{i}-\mathbf{r}_{h}^{j}\right\vert ^{3}}-{{\displaystyle\sum\limits_{k}}
}\frac{c_{2}(\mathbf{r}_{h}^{i}-\mathbf{r}_{e}^{k})}{\left[  (\mathbf{r}%
_{h}^{i}-\mathbf{r}_{e}^{k})^{2}+d^{2}\right]  ^{3/2}},\nonumber
\end{align}
where vectors $\mathbf{r}_{e}^{i}$ and $\mathbf{r}_{h}^{j}$ are
in-plane positions of $i$-th electron and $j$-th hole ($1\leq
i,j\leq N$); $d$ is the interlayer distance. The left-hand side of
Eqs.(\ref{e1}) accounts electron and hole inertia and momentum
damping (mainly due to interaction with acoustic phonons), while the
right-hand side describes the in-layer electron-electron and
hole-hole Coulomb repulsion as well as the interlayer electron-hole
attraction. Hereafter, we normalized time by
$t_{s}=\sqrt{\epsilon}m_{e}^{\ast}\mu_{e}/e$ and all distances by
$r_{s}=\sqrt[3]{m_{e}^{\ast}\mu_{e}^{2}}$, where $e$ and $\mu_{e}$
are electron charge and mobility, respectively, $m_{e}^{\ast}$ is
effective electron mass and $\epsilon$ is dielectric constant of the
layers. The coefficients in Eqs.(\ref{e1}) are
$c_{1}=m_{e}^{\ast}\mu_{e}/\left( m_{h}^{\ast}\mu_{h}\right) $ and
$c_{2}=m_{e}^{\ast}/m_{h}^{\ast}$. To estimate the parameters, we
use well-known experimental values for high-quality undoped
GaAs/AlGaAs QWs \cite{note0} getting $c_{1}\sim1$, $c_{2}\sim0.1$,
$t_{s}\sim10^{-9}$ s, $r_{s}\sim10^{-4}$ cm that gives
$d\sim10^{-2}$ for interlayer spacing of about $10^{-6}$ cm
\cite{But_rev}.

Stationary optical excitation of electrons and holes was modeled by
generating the particles in random positions inside the excitation
spot of radius $r_{0}$ with rate $p$. Initial velocities of
electrons and holes were also chosen randomly within intervals
$(-V_{e}, V_{e})$ and $(-V_{h}, V_{h})$ with ratio
$V_{h}/V_{e}=c_{2}$ to allow the momentum conservation for each e-h
pair created. During the dynamical evolution of the carriers, an
exciton formation happened if an electron and a hole were close
enough to each other, $\left\vert
\mathbf{r}_{e}-\mathbf{r}_{h}\right\vert <a$, where $a(d)$ is a
phenomenological in-layer exciton radius \cite{a_ex}, and their
relative velocity $V=\left\vert
\dot{\mathbf{r}}_{e}-\dot{\mathbf{r}}_{h}\right\vert $ was smaller
than some critical value $V_{c}$ \cite{Vc}. To simplify the
calculations, we did not consider the exciton dynamics and neglected
the interaction of charge carriers with dipole excitons. It means
that as soon as an electron and a hole had formed an exciton, their
dynamics was no longer considered and the position of the formation
event was recorded as a position of photon emission. In addition,
due to the inevitable restrictions in computational power we were
able to simulate $N\lesssim10^{3}$ interacting particles. For this
reason we had to increase the value of $a$.

\textit{Results for 1D case.}---The results of extensive 1D
simulations of Eqs.(\ref{e1}) combined with the exciton formation
condition are shown in Fig.2. The stationary luminescence density
$n_{\rm lum}(x)$ has a large distant peak that correlates with the
ring-shaped pattern observed experimentally
\cite{But_ring,snoke_nat,stern,BLS,But_new}. It is remarkable that
the smaller critical relative velocity $V_{c}$ for the exciton
formation, the larger both the peak amplitude and the distance
between the peak position and the excitation spot $(-x_{0}, x_{0})$
in the center (Fig.2a). Indeed, at extremely large $V_{c}=\infty$,
when an exciton is formed as soon as the instant distance between an
electron and a hole is small enough, the luminescence decreases
monotonically from the excitation spot. For the finite but still
high critical velocities a broad plateau-like maximum appears out of
the centre in the luminescence pattern. Finally, a narrow high peak
of luminescence located on the external plateau edge far away from
the excitation spot develops at lower $V_{c}$. To uncover the origin
of this peak, we plot density distributions for electrons and holes
(Fig.2b). It is seen that the hole distribution is located much
closer to the excitation spot than the electron one. The
luminescence peak forms in the region where electron and hole
densities overlap. In addition, the electron density exhibits
pronounced spatial oscillations beyond the luminescence peak
(Fig.2b). They depend on optical pumping parameters: e.g., the
characteristic period of the oscillations is proportional to
$p^{-1/4}$, where $p$ is pumping rate. By turning off interlayer
electron-hole interaction, we have concluded that the oscillations
originate due to the in-layer electron-electron interaction. In
particular, they can be caused by a kinetic instability of two
interacting electron counter flows: the flow of hot electrons
expanding from the center meets the returning flow of electrons from
the steady electron cloud at the periphery. The back flow can
originate from the depletion of the density of slow electrons in the
luminescence region due to the exciton formation. Nevertheless, a
detailed understanding of the underlying physical mechanism of the
electron density oscillations still requires further studies. The
dependence of the luminescence peak on the electron-hole creation
rate $p$ (inset in Fig.2a) has shown that though the peak height
increases with $p$, its position is almost independent on $p$ for 1D
case. This behaviour differs from that observed experimentally for
2D samples \cite{But_ring}, where the luminescence ring radius grows
essentially with the increase of pumping intensity. The difference
poses a question of modeling the full-value 2D expansion of
photoexcited electron-hole plasma.

\textit{Results for 2D case.}---Due to the quadratic increase of the
number of particles, in 2D case without a boundary (Fig.3) the
stationary state was not reached for the expectation time, though a
clear tendency to slow down was observed for the expansion of the
electron distribution. Remarkably, however, that the luminescence
ring-shaped pattern forms \textit{dynamically} at $t/t_{s}>300$,
fastly expanding afterwards (Fig.3a) (cp. \cite{But_new}). Moreover,
the spatiotemporal dynamics of the 2D electron density (not shown)
has revealed very nontrivial details: at $t/t_{s}\approx300$ the
electron cloud ejects a long "tail" towards the outside of the
excitation area. Due to the corresponding depletion of the internal
part of the cloud, the in-layer separation between electron and hole
distributions becomes more pronounced giving rise to a sharp
luminescence ring in the bilayer plane.

To roughly understand why the dynamical transition occurs in 2D
case, let us reduce Eqs.(\ref{e1}) to the case when the carrier
distributions can be described only by collective ring coordinates
$r_{i}=N^{-1}\int rn_{i}(r)d^{2}r$, where $i=e,h$. In the limit
$r_{e}\gg r_{h}$ the equation of motion for the electron ring reads
\begin{equation}
\ddot{r}_{e}+\dot{r}_{e}\approx\frac{N}{\pi r_{e}^{2}}\ln\left(  2r_{e}%
/\Delta\right)  -\frac{N}{r_{e}^{2}}, \label{3}
\end{equation}
where $\Delta\ll r_{e}$ is a width of the electron ring. (For
simplicity, we neglect the dependences $N\left(t\right)$ and
$\Delta\left(r_{e}\right)$.) The first term in the right-hand side
describes the interaction of electrons within the ring, while the
second term originates from the electron-hole interaction. The
logarithmic factor in the electron-electron interaction term is
absent in 1D case. It results in a fast expansion of the electron
ring for $\ln(2r_{e}/\Delta)>\pi$, i.e., when $r_{e}$ has amounted
to some critical value.

We have confirmed that the principle of the ring pattern formation
in 2D case is the same as that in quasi 1D case: the luminescence
peak appears at the narrow overlap of separated electron and hole
density distributions (Fig.3b). Meanwhile, the electron density
oscillations observed in quasi 1D case are absent in 2D case, even
in the stationary state for the case with a boundary. In the latter
case, we have also observed the onset of the luminescence ring at
high enough pumping rate $p$ (Fig.4, cp. inset in Fig.2a) and the
shift of the ring position to larger $r$ when increasing $p$. The
result correlates well with the experimental dependence (Fig.1a in
\cite{But_ring}) for pumping powers slightly above threshold.

\begin{figure}[ptbh]
\includegraphics[width=8.8cm]{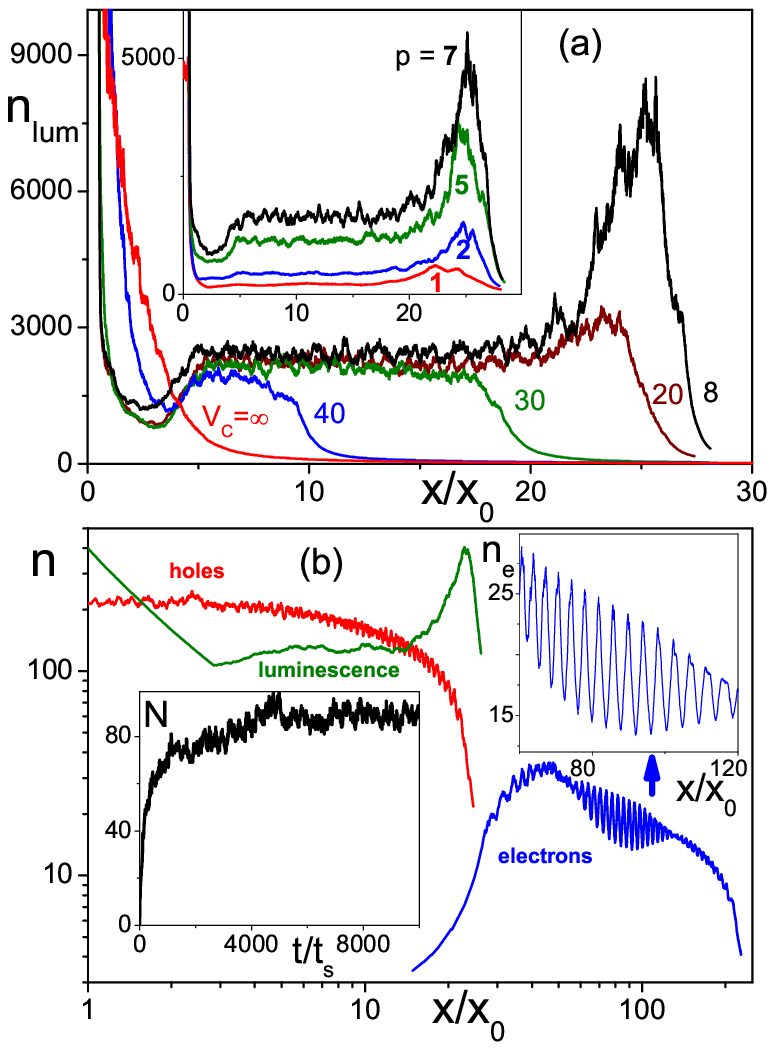}
\caption{\textbf{Ring-shaped luminescence patterns and
spatiotemporal carrier dynamics in quasi 1D case.} \textbf{(a)} The
in-layer distribution of luminescence for different critical
velocities $V_c$ (main panel, $p=10$) and for different pumping
rates $p$ in the excitation spot with size $x_0=1$ (inset,
$V_c=10$). The distant peak of luminescence appears only at
relatively small $V_c$ and the peak intensity grows with increasing
$p$. Constant parameters are $a=0.2$, $c_1=1$, $c_2=0.1$,
$d^2=0.04$, $V_e=50$, integration step $\Delta t=5\cdot 10^{-4}$.
\textbf{(b)} Distributions of electrons (blue curve) and holes (red
curve) for $p=10$ and $V_c=10$ (main panel). The luminescence
distribution exhibits a peak at the overlap of electron and hole
densities which are well separated. Electron density shows clear
oscillations (main panel and top-right inset). In the stationary
state the creation rate of electrons and holes is balanced by the
luminescence rate so the total number $N$ of carriers saturates
(bottom-left inset).} \label{Fig2}
\end{figure}

\begin{figure}[ptbh]
\includegraphics[width=8.8cm]{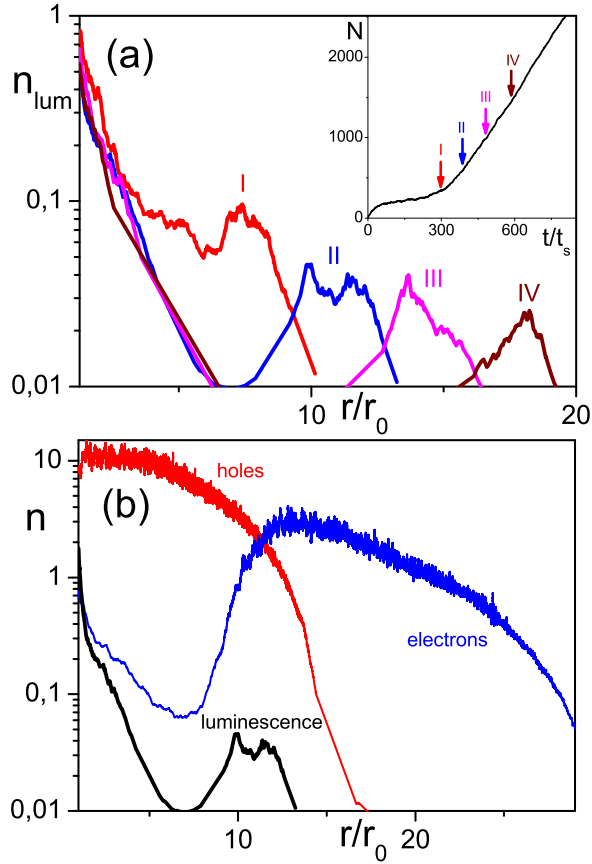}
\caption{\textbf{Spatiotemporal dynamics of the luminescence ring
and electron-hole distributions in 2D case.} All spatial
distributions are averaged over the polar angle. \textbf{(a)}
Snapshots of luminescence distribution $n_{\rm lum}(r,t)$ averaged
over time window ($t_1/t_s$, $t_2/t_s$): (300, 400) for the red (I)
curve, (400, 500) for the blue (II) curve, (500, 600) for the
magenta (III) curve, (600, 700) for the brown (IV) curve. The peak
of $n_{\rm lum}$ mimics the ring-like patters observed in
experiments \cite{But_ring,snoke_nat,BLS,stern,But_new}. In some
time after the pumping is switched on, the ring forms and expands
even though the stationary regime is not reached (inset). Constant
parameters: $a=1.5$, excitation spot size $r_{0}=4$, $p=10$,
$c_{1}=1$, $c_{2}=0.25$, $d^{2}=2.25$, $V_e=50$, $V_c=10$, $\Delta
t=5\cdot 10^{-4}$. \textbf{(b)} The snapshots of electron and hole
densities as well as luminescence pattern averaged over time window
(400, 500). The ring pattern forms at the overlap of legibly
separated electron and hole distributions.}
\label{Fig3}%
\end{figure}

\begin{figure}[ptbh]
\includegraphics[width=8.8cm]{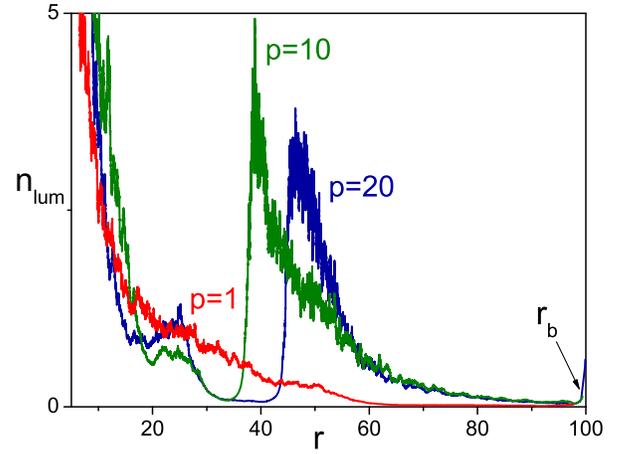}
\caption{\textbf{The luminescence ring at different values of
pumping rate $p$ for 2D case with boundary (mirror reflection)
$r_{b}=100$ in the stationary state.} All other parameters are the
same as for Fig.3.}
\end{figure}

\textit{Qualitative picture.}---The macroscopic electron-hole
spatial separation and luminescence ring-shaped pattern formation
can be explained as follows. At high photoexcitation power the
carrier densities in the excitation spot quickly reach their
critical values ($\sim d^{-2}$) when the repulsive in-layer Coulomb
forces between particles become stronger than the attractive
interlayer force. It leads to the appearance of in-layer electric
fields ejecting the "surplus" electrons and holes from the
excitation spot region. Then the e-h densities in the spot grow to
the critical values again and the ejection process recurs. Due to
very high mobilities of the carriers in GaAs quantum wells at low
temperature \cite{mob0,mob,cp}, the initial velocities of the
ejected particles are also high. The absolute values of the
velocities are restricted by the emission threshold of an optical
phonon that gives $v_{\max}\sim10^{7}$ cm/s \cite{note}. In turn,
the velocity ratio for the hot electrons and holes has the same
order of magnitude as the ratio between their mobilities
($\mu_{e}/\mu_{h}\gtrsim10$ for GaAs/AlGaAs QWs at temperatures
$T\sim1$ K \cite{mob}.) Due to this, the electron distribution is
mainly located much farther from the excitation spot than the hole
one. During the in-plane expansion the electrons and holes cool down
emitting acoustic phonons. In addition, their relative velocities
decrease due to the interlayer Coulomb drag \cite{screen}. Finally,
the luminescence ring appears in the overlapping region between the
stationary densities of the slowed electrons and holes. Note that at
typical electron-acoustic phonon scattering time
$\tau_{e-ac}\sim10^{-9}$ s \cite{phon} one comes to the promising
estimate for the ring radius $R\sim v_{\max}\tau_{e-ac}\sim0.1$ mm
(cp. \cite{But_ring,cp}).

\textit{Conclusion.}---We have simulated a spatiotemporal evolution
of the locally photoexcited electrons and holes localized in two
separate layers. In contrast to previous theoretical studies
\cite{BLS,theor,Ha,But_new}, we have shown that both in-plane
spatial separation of electrons and holes and the ring-shaped
luminescence pattern are formed essentially due to the in-layer
Coulomb repulsion. A specific reason for the separation is found to
be the difference in electron and hole mobilities. In turn, a
specific reason for the luminescence ring is the dependence of
exciton formation probability on the electron-hole relative
velocity. We also observed drastically different electron and hole
dynamics in 1D and 2D cases and, more importantly, pronounced
stationary spatial oscillations of the electron density in 1D case.

We thank A. S. Alexandrov, Yu. M. Kagan and F. V. Kusmartsev for
helpful discussions.


\begin{thebibliography}{30}
\bibitem {note0}We take typical $\mu_{e}\sim10^{7}$ cm$^{2}$/(Vs)
\cite{mob0} for temperatures $T\lesssim1$ K and electron densities
$n_{e}\sim10^{11}\div10^{12}$ cm$^{-2}$ (excitation spot),
$\epsilon=12.8$ for GaAs (see also \cite{note}) and
$\mu_{h}\sim0.1\mu_{e}$ \cite{mob}.
\bibitem {a_ex}Due to the finite interlayer distance $d$, the exciton radius $a(d)$
can be essentially larger than the Bohr radius
$a_{B}=\epsilon\hbar^{2}/me^{2}$ of a purely 2D exciton
($a_{B}\approx10^{-6}$ cm  for GaAs).
\bibitem {Vc}There are two critical relative velocities so one should
choose the minimal value. One appears due to the finite interlayer
distance $d$, $V_{c1}=\hbar/\left(  md\right)$, where $m$ is e-h
reduced mass. If $V>V_{c1}$ the Fourier transform of interlayer
Colulomb interaction potential drops as $\exp(-V/V_{c1})$
\cite{screen}. At $d\approx10^{-6}$ cm \cite{But_rev} and
$m\approx0.06m_{e}$ in GaAs, $V_{c1}\sim10^{7}$ cm/s. Another comes
from the exciton formation condition: electron-hole scattering rate
should not exceed carrier-acoustic phonon one, $V<V_{c2}$. Its
estimate is $V_{c2}\sim 1/\sqrt{n\tau^{2}}$,
$\tau=\min(\tau_{e-ac},\tau_{h-ac})$ and $n=\max (n_{e},n_{h})$. For
$\tau\sim10^{-9}$ s \cite{phon} and $n\sim10^{10}$ cm$^{-2}$
\cite{But_rev} one gets $V_{c2}\sim10^{4}$ cm/s.
\bibitem {note}To contribute to the transport, the value of photoexcited carrier
velocity should be restricted by optical phonon emission,
$m^{\ast}v^{2}/2\leq \hbar\omega_{LO}$ , where for GaAs
$\hbar\omega_{LO}\approx37$ $meV$ is the optical phonon energy, and
effective electron and heavy hole masses are
$m_{e}^{\ast}\approx0.067m_{e}$ and $m_{hh}^{\ast}\approx0.5m_{e}$
($m_{e}$ is bare electron mass). The estimates result in
$v_{\max}\sim10^{7}$ cm/s. The corresponding de Broglie wavelength
of the carriers $\lambda_{dB}\sim10^{-6}$ cm.
\end{thebibliography}
\end{document}


\title{Erratum: Ring-shaped luminescence patterns in a locally photoexcited electron-hole bilayer [Phys. Rev. B 81, 193403 (2010)]}

\author{A. V. Paraskevov, S. E. Savel'ev}

\noaffiliation

\pacs{71.35.--y, 73.50.Gr, 73.63.-b, 78.67.De}

\maketitle

We have found a typo in the Y-projection of the interlayer
electron-hole force in the MD code for 2D case. It affects the
results shown in Fig.3 and Fig.4 of our original paper. Due to the
typo the angular symmetry of carrier spatial distributions was
broken that led to the absence of balance between pumping and decay
in the 2D case without boundary and, eventually, to the permanent
growth of the total number $N$ of carriers in time $t$ (inset in
Fig.3a).

Simulations by the revised code have shown that there is a
saturation of $N(t)$ in the 2D case without boundary (see inset in
Fig.1 here). The balance occurs at about 350 particles that
corresponds to the first plateau of $N(t)$ shown on the inset in
Fig.3a. The spatial distributions of carriers and luminescence are
quite similar to that shown in Fig.3b, i.e., there is clear spatial
separation of the electron and hole distributions though the
luminescence ring is less pronounced (main panel in Fig.1 here).

The dependence of the ring position on the pumping rate $p$ \cite{1}
becomes very weak: the increase of $p$ in few times leads to the
growth of the ring radius on few percents \cite{2}. In this
relation, the results shown in Fig.4 are erroneous. Note that at
$p=10$ for the 2D case without boundary the maximal radius of the
electron distribution is about 100 so the boundary $r_{b}=100$ plays
no role.

\begin{figure}[htbp]
\includegraphics[width=9.5cm]{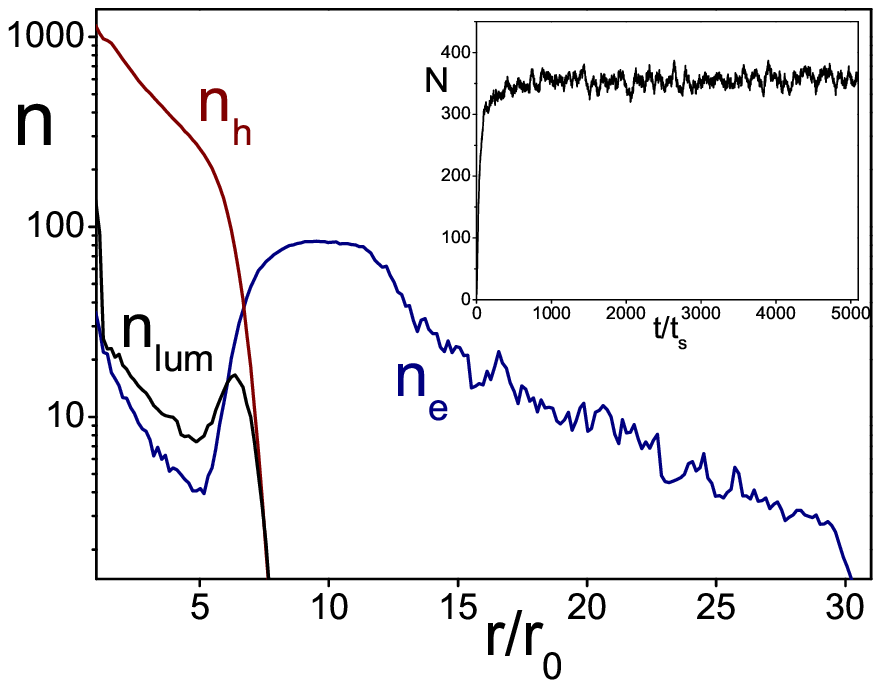}
\caption{Stationary spatial distributions of electrons ($n_{e}$),
holes ($n_{h}$) and luminescence ($n_{\rm lum}$) averaged over time
window (500, 5000) in the 2D case without boundary. Inset:
dependence $N(t)$. All parameters are the same as for Fig.3.}
\end{figure}

So, the main consequence of the revised code is the saturation in
the 2D case without boundary. Note that it brings an exciting
opportunity to perform non-time-consuming molecular-dynamic
simulations of spatiotemporal dynamics of charge carriers in
mesoscopic 2D systems.